\documentclass[british,structabstract]{aa}
\usepackage{mathptmx}
\usepackage[T1]{fontenc}
\usepackage{url}
\usepackage{graphicx}

\makeatletter

\providecommand{\tabularnewline}{\\}

\bibpunct{(}{)}{;}{a}{}{,} 
\usepackage{hyperref}
\usepackage{longtable}
\usepackage{lscape}
\usepackage{txfonts}

\newcommand{\taur}{\tau_{\mathrm{Ross}}}

\newcommand{\hav}{\left\langle \mathrm{3D}\right\rangle}
\newcommand{\havz}{\hav_z}

\titlerunning{Surface-effect corrections for the solar model}
\authorrunning{Z. Magic \& A. Weiss}

\@ifundefined{showcaptionsetup}{}{%
 \PassOptionsToPackage{caption=false}{subfig}}
\usepackage{subfig}
\makeatother

\usepackage{babel}
\begin{document}

\title{Surface-effect corrections for the solar model}

\author{Z. Magic\inst{1,2} and A. Weiss\inst{3}}

\institute{Niels Bohr Institute, University of Copenhagen, Juliane Maries Vej
30, DK--2100 Copenhagen, Denmark \and  Centre for Star and Planet
Formation, Natural History Museum of Denmark, {\O}ster Voldgade 5-7,
DK--1350 Copenhagen, Denmark\and Max-Planck-Institut für Astrophysik,
Karl-Schwarzschild-Str. 1, 85741 Garching, Germany\\
\email{magic@nbi.dk}}

\date{Received ...; Accepted...}

\abstract{Solar p-mode oscillations exhibit a systematic offset towards higher
frequencies due to shortcomings in the 1D stellar structure models,
in particular, the lack of turbulent pressure in the superadiabatic
layers just below the optical surface, arising from the convective
velocity field. }
{We study the influence of the turbulent expansion, chemical composition,
and magnetic fields on the stratification in the upper layers of the
solar models in comparison with solar observations. Furthermore, we
test alternative $\hav$ averages for improved results on the oscillation
frequencies. }
{We appended temporally and spatially averaged $\hav$ stratifications
to 1D models to compute adiabatic oscillation frequencies that we
then tested against solar observations. We also developed depth-dependent
corrections for the solar 1D model, for which we expanded the geometrical
depth to match the pressure stratification of the solar $\hav$ model,
and we reduced the density that is caused by the turbulent pressure.
}
{We obtain the same results with our $\hav$ models as have been reported
previously. Our depth-dependent corrected 1D models match the observations
to almost a similar extent as the $\hav$ model. We find that correcting
for the expansion of the geometrical depth and the reducing of the
density are both equally necessary. Interestingly, the influence of
the adiabatic exponent $\Gamma_{1}$ is less pronounced than anticipated.
The turbulent elevation directly from the $\hav$ model does not match
the observations properly. Considering different reference depth scales
for the $\hav$ averaging leads to very similar frequencies. Solar
models with high metal abundances in their initial chemical composition
match the low-frequency part much better. We find a linear relation
between the p-mode frequency shift and the vertical magnetic field
strength with $\delta v_{nl}=26.21B_{z}\,\left[\mu\mathrm{Hz}/\mathrm{kG}\right]$,
which is able to render the solar activity cycles correctly. }
{}


\maketitle

\section{Introduction\label{sec:Introduction}}

The Sun shows p-mode oscillations that are due to the stochastic excitation
by turbulent convection. Certain frequencies are damped and others
are amplified, which leads to a characteristic set of frequencies
observed on the Sun. In helioseismology, observed oscillation frequencies
are compared with theoretical model predictions to probe the solar
structure below the optical surface \citep{christensen-dalsgaard_helioseismology_2002}.
In 1D stellar structure calculations, convection is modelled with
the mixing length theory \citep{bohm-vitense_uber_1958}, hence the
mismatch in the superadiabatic region (SAR) just below the optical
surface leads to systematic residuals for the high-frequency modes.
These are termed surface effects. To account correctly for convection,
and in particular, for the turbulent pressure, we need to employ 3D
hydrodynamic models, where convection emerges from first principles
\citep{stein_simulations_1998}. \citet{rosenthal_convective_1999}
found that appending a mean stratification from a 3D model \citep{stein_simulations_1998}
to a 1D solar model \citep{1996Sci...272.1286C} can indeed reduce
the surface effects by providing additional support through turbulent
pressure and a concomitant expansion of the acoustic cavity at the
surface by 150 km, which in turn reduces the oscillation frequencies.
\citet{sonoi_surface-effect_2015} calibrated surface-effect corrections
for other stars than the Sun from 3D models. In the present work,
we repeat the efforts of \citet{rosenthal_convective_1999} with the
solar model from the \textsc{Stagger grid} \citep{magic_stagger-grid:_2013-1}
and study the solar surface effects in detail.

\section{Methods\label{sec:Methods}}

\subsection{Theoretical models\label{sub:Models}}

\begin{figure}
\includegraphics[width=88mm]{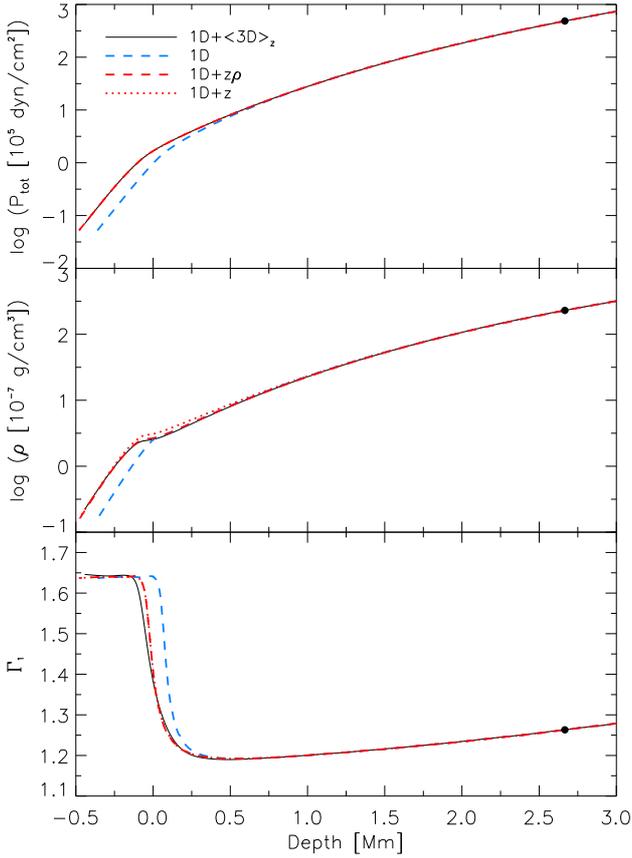}

\caption{\label{fig:sun_structure}Total pressure, density, and adiabatic exponent
for the 1D model combined with the $\hav_{z}$ model. We marked the
location of the transition between $\hav_{z}$ and 1D (\emph{black
dot}). In additition to the original 1D model (\emph{blue dashed line}),
we also show the values of the 1D model corrected for both the geometrical
depth and density (\emph{red dashed line}) and corrected for the depth
alone (\emph{red dotted line}).}
\end{figure}
\begin{figure}
\includegraphics[width=88mm]{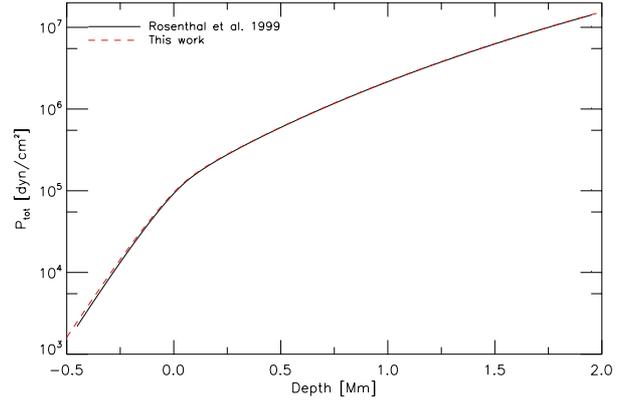}\caption{\label{fig:Rosenthal_pressure}Total pressure vs. depth averaged from
the 3D models of this work and compared with \citet{rosenthal_convective_1999}.}

\end{figure}
\begin{figure}
\includegraphics[width=88mm]{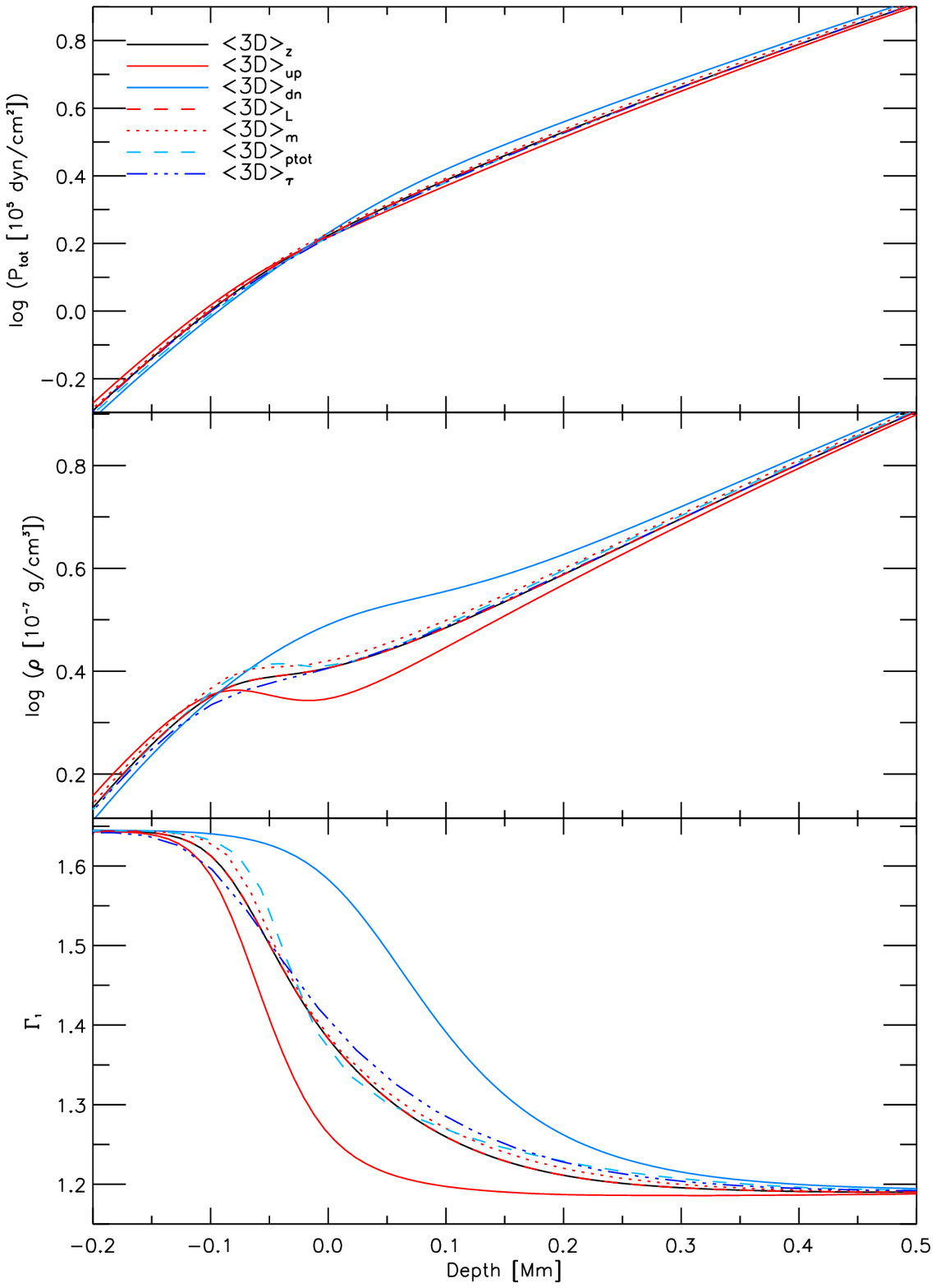}\caption{\label{fig:sun_averages}Similar to Fig. \ref{fig:sun_structure},
but showing models with different $\hav$ averages. The geometrically
averaged model, $z$ (solid black line), is identical to the model
shown in Fig. \ref{fig:sun_structure} (black solid line). Inward
of a depth of roughly -0.1 Mm, the averages of the upflows, $\hav_{\mathrm{up}}$,
are the lowest curves, while the downflows, $\hav_{\mathrm{dn}}$,
are the uppermost curves in all three panels.}
\end{figure}
We considered the solar mean $\hav$ model taken from the \textsc{Stagger
grid} \citep[see][for more details]{magic_stagger-grid:_2013}, which
was computed with the \textsc{Stagger} code. In the 3D models, the
equation-of-state (EOS) from \citet{1988ApJ...331..815M} is used,
and the opacities are taken from the \textsc{marcs} package \citep{2008A&A...486..951G}.
The numerical resolution is $240^{3}$ and twelve opacity bins were
considered for the radiative transfer. 

For the 1D model we took a standard solar calibrated model (including
atomic diffusion) computed with the \textsc{garstec} stellar evolution
code \citep[see][for more details]{weiss_garstecgarching_2008,magic_using_2010}.
In the 1D model, we used the EOS by OPAL 2005 \citep{1996ApJ...456..902R}
for higher temperatures, while for the lower temperatures we extended
the EOS by those from \citet{1988ApJ...331..815M}, for consistency
between the 3D and 1D models. The opacity in the \textsc{garstec}
model is a combination of results from OPAL \citep{1996ApJ...464..943I}
and \citet{2005ApJ...623..585F} for higher and lower temperatures,
respectively. In all models, we assumed the solar chemical composition
by \citet{asplund_chemical_2009}, except for the oscillation frequencies
in Sect. \ref{sub:Abundances}.

The pressure stratification of our $\hav$ model shown in Fig. \ref{fig:sun_structure}
is very similar to the gas gamma model by \citet{rosenthal_convective_1999}.
Our 3D model exhibits a slightly higher pressure stratification only
above the optical surface (see Fig. \ref{fig:Rosenthal_pressure}).
This is probably owed to the differences in the input physics, the
treatment of the radiative transfer, and numerical resolution (theirs
is lower with $100^{2}\times82$). When we compared our solar model
with models with higher numerical resolution ($480^{3}$ and $960^{2}\times480$),
we found the differences in the stratifications of the models to be
minute. Comparisons with other numerical codes (\textsc{co5bold} and
\textsc{muram}) also result in very similar temperature and pressure
stratifications \citep[see Figs. 4 and 5 in][]{2012A&A...539A.121B}.
These comparisons mean that our temperature and pressure stratifications
are reliable.

To compute the theoretical eigenmode frequencies, we employed the
adiabatic oscillation code \textsc{adipls} \citep{christensen-dalsgaard_adiplsaarhus_2008}.
Here, the independent variables are the total pressure, $p_{\mathrm{tot}}$,
the density, $\rho$, and the adiabatic exponent, $\Gamma_{1}$, (see
Fig. \ref{fig:sun_structure}). \citet{rosenthal_convective_1999}
computed oscillation frequencies with the spatially and temporally
averaged adiabatic exponent, to which they referred as the ``gas
gamma model''. They also worked with a ``reduced gamma model''.
Based on the differences between the gas gamma and the reduced gamma
model, they found a difference of $\sim10\,\mu\mathrm{Hz}$ at higher
frequencies. However, their reduced gamma model disagreed with observations
because the adiabatic exponent was too low. Therefore, we used the
spatial and temporal averages of $\Gamma_{1}$ in the present study
(gas gamma model). We also computed the reduced gamma, $\Gamma_{1}^{\mathrm{red}}=\left<p_{\mathrm{gas}}\Gamma_{1}\right>/\left<p_{\mathrm{gas}}\right>$,
but we found only very little difference to the plain values of $\Gamma_{1}$.
Since the surface effects are less dependent on the angular degree,
we considered the lowest mode ($l=0$) in comparison with the observational
data provided by GOLF measurements \citep{1997SoPh..175..227L}.

\subsection{Temporal and spatial averaging\label{sub:Averaging}}

\begin{table}
\caption{\label{tab:mean_3d}Overview of the different mean $\hav$ models
and their reference depth scale for the horizontal or spatial averaging.}
\begin{tabular}{ll}
\hline 
Symbol & Spatial averages over layers of constant \tabularnewline
\hline 
\hline 
$\havz$ & geometrical depth\tabularnewline
$\hav_{\mathrm{up}}$ & geometrical depth for the upflows\tabularnewline
$\hav_{\mathrm{dn}}$ & geometrical depth for the downflows\tabularnewline
$\hav_{L}$ & (pseudo-Lagrangian) geometrical depth \tabularnewline
$\hav_{m}$ & column mass density\tabularnewline
$\hav_{p_{\mathrm{tot}}}$ & total pressure\tabularnewline
$\hav_{\tau}$ & acoustic depth\tabularnewline
\hline 
\end{tabular}
\end{table}
To compute the spatially and temporally averaged mean $\hav$ is nontrivial,
as we showed in \citet{magic_stagger-grid:_2013-1} for spectroscopy.
For the application in helio- and asteroseismology, we similarly needed
to test and find a reference depth-scale to average independent variables
in the most suitable way. Therefore, we considered seven different
averages (see Table \ref{tab:mean_3d}) that we explain in the following.
The geometrical average (denoted with $z$) is used by default, since
it fulfils the hydrostatic equilibrium. We computed geometrical averages
separated into up- and downflows (denoted with $\mathrm{up}$ and
$\mathrm{dn}$), based on the sign of the vertical velocity component.
Furthermore, we also used pseudo-Lagrangian averages (denoted with
$L$) that were spatially averaged over geometrical depth and temporally
averaged by mapping to a fixed column mass scale to remove the contribution
of the p-mode oscillations from the spatial averages \citep[see][]{2014MNRAS.442..805T}.
Optical depth can be ruled out as a reference depth scale, because
the averages are then correlated to the temperature \citep[see][for more details]{magic_stagger-grid:_2013}.
We considered averages over layers of constant column mass density
\begin{eqnarray}
m & = & \int\rho dz,
\end{eqnarray}
acoustic depth 
\begin{eqnarray}
\tau & = & \int\frac{1}{c_{s}}dz,
\end{eqnarray}
and total pressure $p_{\mathrm{tot}}$. In the photosphere, where
the fluctuations are strongest, we can expect differences between
the different averages that might alter the oscillation frequencies.
In Fig. \ref{fig:sun_averages} we show the total pressure, density,
and adiabatic exponent for the different $\hav$ averages. The different
averages exhibit small differences for the independent variables,
except for the averages for the up- and downflows, which show larger
differences than the other averages. In particular, this is the case
for the density and adiabatic exponent, while the total pressure shows
only smaller differences. The averages of the upflows depict the hot,
ascending granules, while the downflows depict the cold, descending
intergranular lane at the optical surface. The pseudo-Lagrangian averages
(red dashed lines) are indistinguishable from the geometrical averages
(solid black lines). As mentioned above the pseudo-Lagrangian averages
are spatially averaged over geometrical depth and differ only in the
temporal averaging, which seems to have very little effect on the
stratification. Furthermore, we note that the averages over the column
mass density are clearly distinct from the former because they employ
the column mass density as the reference depth scale.

\subsection{Including the effects of the turbulent expansion\label{sub:Including-the-effects}}

The 3D models are very shallow, therefore we need to append the $\hav$
models to a 1D model to study the effect on the p-mode oscillations.
To do this, we employed two methods (Sects. \ref{sub:Appending-3d}
and \ref{sub:Depth-dependent-corrections}) to include the effects
of the turbulent expansion on the 1D model. The first method is similar
to the one by \citet{rosenthal_convective_1999}, and we appended
a $\hav$ model to a 1D model. In the second method, we expanded the
geometrical depth in the top layers of the 1D model to achieve the
same total pressure stratification as for the $\hav$ model.

\subsubsection{Appending the $\hav$ on the 1D model \label{sub:Appending-3d}}

We appended the $\hav_{z}$ model averaged over layers of constant
geometrical depth, where the zero-point was set to coincide with the
optical surface, that is, $\left\langle \taur\right\rangle =0$, to
the 1D model at the bottom of the $\hav_{z}$ model. To do this, we
considered the total pressure in the deepest layers of the $\hav_{z}$
model, $p_{\mathrm{tot}}^{\mathrm{bot}}$, and matched the difference
in geometrical depth
\begin{eqnarray}
\Delta z^{\mathrm{bot}} & = & z^{\mathrm{3D}}(p_{\mathrm{tot}}^{\mathrm{bot}})-z^{\mathrm{1D}}(p_{\mathrm{tot}}^{\mathrm{bot}}),\label{eq:bot_shift}
\end{eqnarray}
which is a single depth-independent correction value. We obtained
for $\Delta z^{\mathrm{bot}}$ a shift of 81.6 km. Next, we shifted
the geometrical depth of the $\hav_{z}$ models for the difference,
that is, $z^{\mathrm{3D,shifted}}=z^{\mathrm{3D}}+\Delta z^{\mathrm{bot}}$.
Finally, we interpolated the independent variables $p_{\mathrm{tot}}$,
$\rho$, $\Gamma_{1}$ from the $\hav_{z}$ to the geometrical depth
of the 1D model ($z^{\mathrm{1D}}$) and merged both models into a
single model. The appended model is expanded relative to the 1D model
and the stratifications of $p_{\mathrm{tot}}$, $\rho$ and $\Gamma_{1}$
are visibly modified at the top (see Fig. \ref{fig:sun_structure}).
The global radius is increased by the total elevation at the surface
with $\Delta z\left(0\right)=105\,\mathrm{km}$, as determined in
Sect. \ref{sub:Depth-dependent-corrections}. This value can be compared
to the findings by \citet{rosenthal_convective_1999} of 150 km. The
correction of the depth at the bottom of the $\hav$ model, $\Delta z^{\mathrm{bot}}=81.6\,\mathrm{km}$,
is necessary, since the 1D model compensates the missing turbulent
pressure and the higher temperatures at the surface of the 3D model
with a higher pressure scale height below the optical surface. The
different temperature stratification in the 3D model is a result of
the temperature sensitivity of the opacity and inhomogeneities of
the granulation \citep[see Sect. 5.2 in][]{rosenthal_convective_1999}.
On the other hand, without the depth correction, the 3D and 1D model
would exhibit a discontinuous transition at the connection point.
Both of the latter are important to improve the oscillation frequency,
as we show below. We refer to the $\hav_{z}$ model appended to 1D
model as the 3D model.

\subsubsection{Depth-dependent structure corrections\label{sub:Depth-dependent-corrections}}

\begin{figure}
\includegraphics[width=88mm]{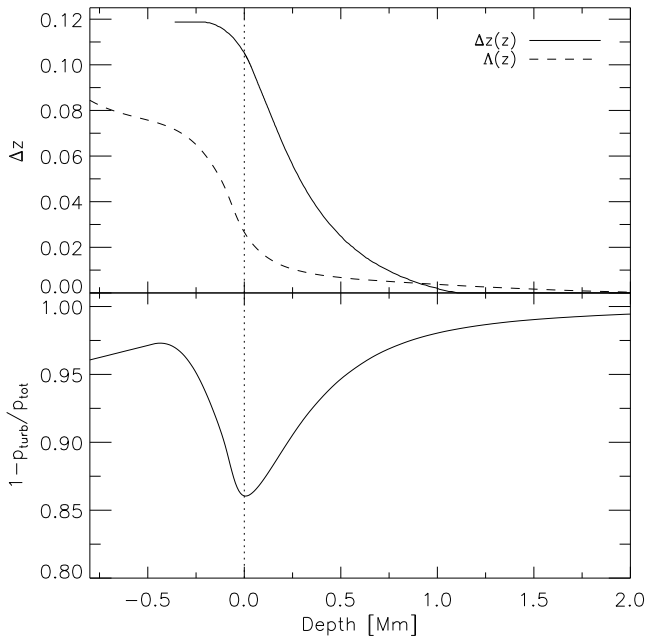}

\caption{\label{fig:corrections}Depth-dependent corrections for the geometrical
depth, $\Delta z$, and corrections of the density, $1-p_{\mathrm{turb}}/p_{\mathrm{tot}}$,
in the 1D solar model (top and bottom panel, respectively). Furthermore,
we show the turbulent elevation from Eq. \ref{eq:turb_levitation}
(\emph{dashed line} in the top panel). The location of the optical
surface is indicated (\emph{vertical dotted line}).}
\end{figure}

Instead of appending a $\hav$ structure to the 1D model, we employed
depth and density corrections directly on the 1D stellar structure
as well. The idea behind this is very simple: we wish to find a depth
correction that expands the geometrical depth, so that the same pressure-depth
relation of the 3D model is realised for the 1D model. To do this,
we determined the difference in geometrical depth, which is necessary
to yield the same pressure stratification. The depth-dependent correction
factor is determined by
\begin{eqnarray}
\Delta z\left(z\right) & = & z^{\mathrm{3D}}(p_{\mathrm{tot}}^{\mathrm{3D}})-z^{\mathrm{1D}}(p_{\mathrm{tot}}^{\mathrm{1D}}),\label{eq:depth_correction}
\end{eqnarray}
which can expand the geometrical depth of the 1D model to the match
the total pressure stratification of the $\hav$ model. We note that
the geometrical depth of the 3D model, $z^{\mathrm{3D}}$, in Eq.
\ref{eq:depth_correction} is corrected for the depth shift at the
bottom with Eq. \ref{eq:bot_shift}, that is, $z^{\mathrm{3D,shifted}}=z^{\mathrm{3D}}+\Delta z^{\mathrm{bot}}$.
Furthermore, we note that in contrast to Eq. \ref{eq:bot_shift},
here $\Delta z\left(z\right)$ is depth dependent. We neglected negative
depth corrections below the optical surface for consistency. Then,
the resulting correction was applied to the geometrical depth, that
is, $z^{*}=z+\Delta z\left(z\right)$, which leads to an expansion
of the entire stellar structure at the top alone. This results in
an extension of the solar radius at the optical surface, meaning that
from Eq. \ref{eq:depth_correction} we obtain $\Delta z\left(0\right)=105\,\mathrm{km}$
(see Fig. \ref{fig:corrections}).

The corrected geometrical depth deviates from the hydrostatic equilibrium
in the SAR, therefore, we corrected the density stratification by
reducing it by the ratio of the turbulent pressure and the total pressure:
\begin{eqnarray}
\rho^{*} & = & \rho\left(1-p_{\mathrm{turb}}/p_{\mathrm{tot}}\right),\label{eq:density_correction}
\end{eqnarray}
since the missing hydrostatic support from the $p_{\mathrm{turb}}$
is balanced by higher densities in the 1D model. On the other hand,
we can also determine the density stratification, which fulfils the
hydrostatic equilibrium. Then, this correction leads to the identical
density stratification as given by the $\hav$ model, since the corrected
geometrical depth results in an identical pressure stratification
with depth by construction, and the equation of hydrostatic equilibrium
(Eq. \ref{eq:hydrostatic_equilibrium}) is only fulfilled by a unique
density stratification.

\begin{figure*}
\subfloat[\label{fig:3d_vs_1d}]{\includegraphics[width=88mm]{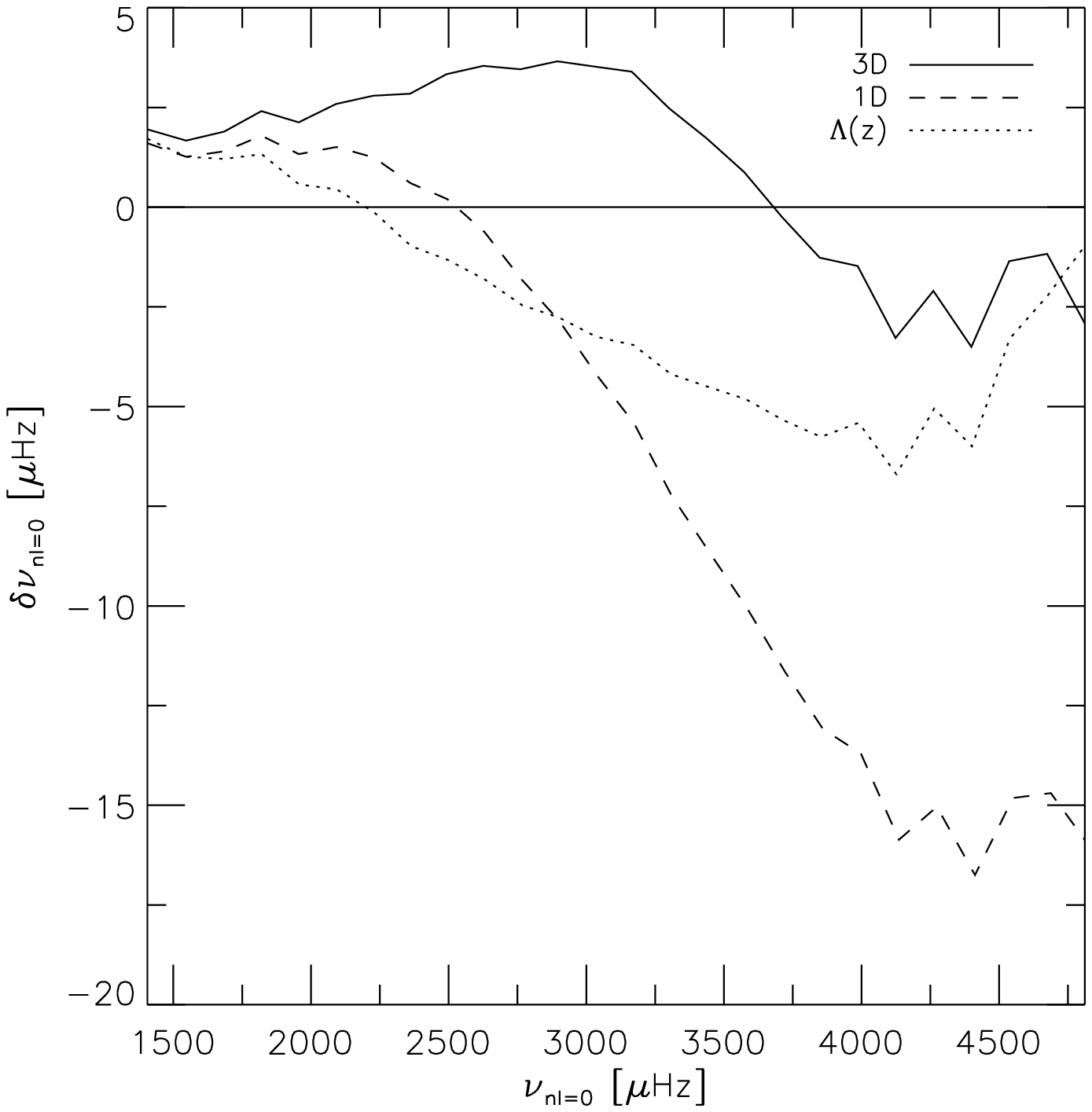}

}\subfloat[\label{fig:rosenthal6}]{\includegraphics[width=88mm]{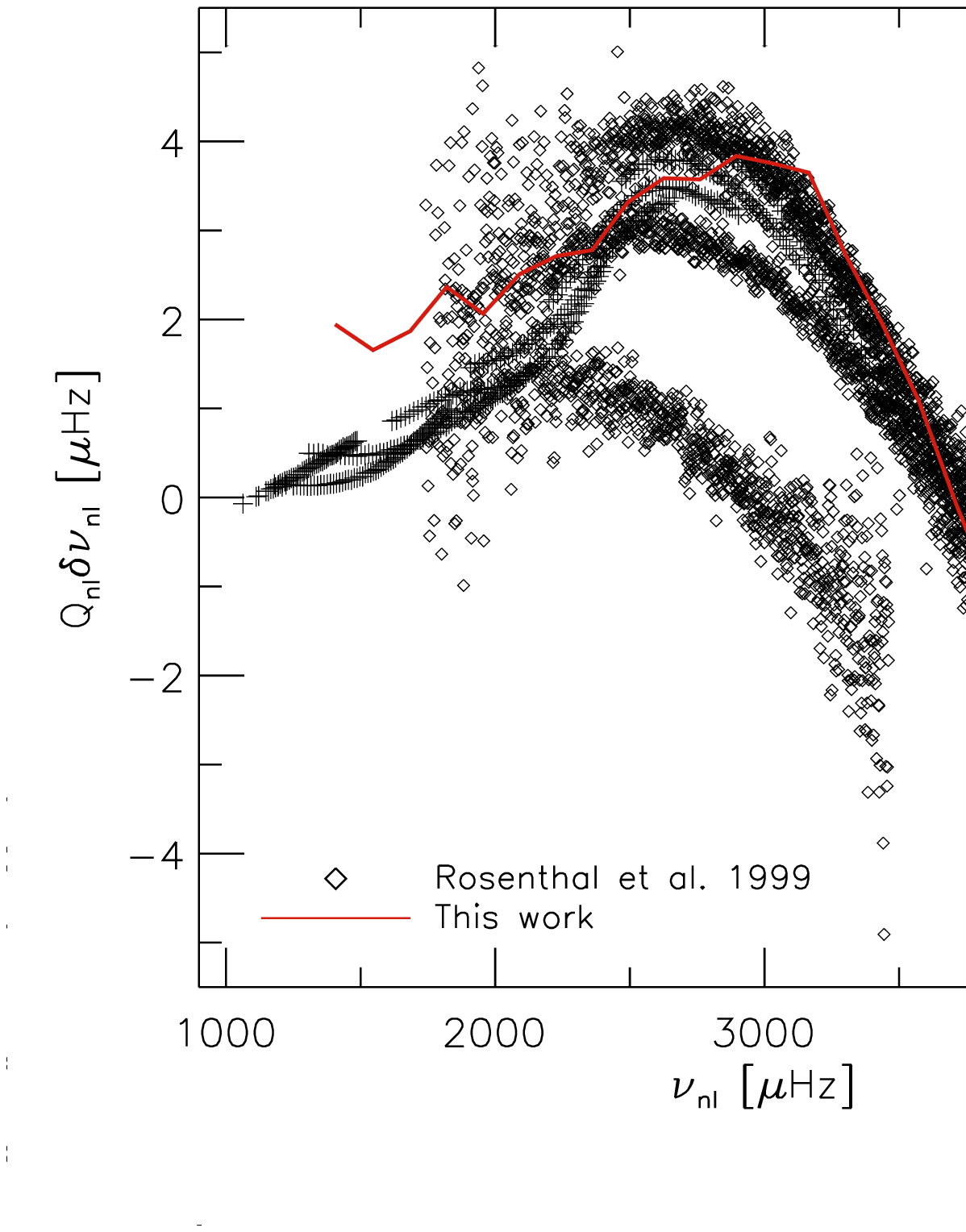}

}

\caption{\textbf{(a)} Frequency residuals vs. frequency for two solar models
in comparison with solar observations for a low-degree mode with $l=0$.
The differences are retrieved by $\delta\nu_{nl}=\nu_{nl}^{\mathrm{obs}}-\nu_{nl}^{\mathrm{mod}}$.
Shown are the 3D corrected model (\emph{black solid line}), the standard
1D solar model (\emph{dashed lines}) and a 1D model expanded with
the turbulent elevation (\emph{dotted lines}; see Eq. \ref{eq:turb_levitation}).
The height of zero differences is indicated (\emph{horizontal solid
line}). \textbf{(b)} We also show the residuals for different modes
scaled with $Q_{nl}$ resulting from a solar model corrected with
3D model (GGM: gas gamma model) published by \citet{rosenthal_convective_1999}
(black symbols) in comparison with our result for $l=0$ (red solid
line).}
\end{figure*}
In Fig. \ref{fig:corrections} we show both the depth and density
correction. The corrected 1D model including $z^{*}$ and $\rho^{*}$
is also shown in Fig. \ref{fig:sun_structure} (red dashed lines),
which is very close to the $\hav$ stratification. The density stratification
without the correction (Eq. \ref{eq:density_correction}) is only
slightly higher, and only minor differences are visible for $\Gamma_{1}$.
As we show in Sect. \ref{sub:Corrected-structures}, both have effects
of very different magnitudes on the oscillation frequencies.

The hydrostatic equilibrium can be retrieved by averaging the momentum
equation, which results in
\begin{eqnarray}
\frac{d(\bar{p}_{\mathrm{th}}+\bar{\rho}\bar{u}_{z}^{2})}{dz} & = & \bar{\rho}g_{z},\label{eq:hydrostatic_equilibrium}
\end{eqnarray}
with $\bar{u}_{z}$ being the density-weighted average of the vertical
velocity component (often referred to as turbulent velocity) and $p_{\mathrm{turb}}=\rho u_{z}^{2}$
the turbulent pressure, which is an additional pressure support against
gravity. The density correction stems mostly from the vertical component
of the velocity field. By separating the support from $p_{\mathrm{turb}}$
and integrating for the depth, we obtain the turbulent elevation 
\begin{eqnarray}
\Lambda\left(z\right) & = & -\int_{-\infty}^{z}\frac{dp_{\mathrm{turb}}}{\rho g},\label{eq:turb_levitation}
\end{eqnarray}
which depicts the expansion caused by the turbulent velocity field
that is due to convection in SAR \citep{1997MsT..........3T}. In
Fig. \ref{fig:corrections} we also show the turbulent elevation (dashed
line). This has a shape similar to $\Delta z\left(z\right)$, but
the amplitude of the total elevation at the surface is mismatched
with $\Lambda\left(0\right)=26\,\mathrm{km}$, in contrast to $\Delta z\left(0\right)=105\,\mathrm{km}$.
This difference propagates into a larger difference in oscillation
frequencies, rendering the use $\Lambda\left(z\right)$ directly as
futile, meaning that the resulting frequencies are unable to match
the observations (see Fig. \ref{fig:3d_vs_1d}). 

To append $\hav$ models to 1D structures is straightforward in the
case of solar parameters; for other stars, such as red giants or turn-off
stars, however, this becomes more difficult, because the gradients
of 3D and 1D models often do not match. A possible solution of this
dilemma is to include the missing turbulent pressure in the 1D model.
To do this, we can use the ratio between the turbulent and total pressure,
which quickly drops close to zero below the SAR. We corrected for
the density first with Eq. \ref{eq:density_correction} and for the
ratio in pressures from the $\hav$ model. Then, we integrated for
the geometrical depth under hydrostatic equilibrium with Eq. \ref{eq:hydrostatic_equilibrium},
which expanded the geometrical depth. To obtain good results, we needed
to apply a constant factor of $3/2$ to the ratio of pressures in
Eq. \ref{eq:density_correction}. We emphasize that the advantage
of this approach is that the $\hav$ model does not need to be matched
at the bottom, which results in a more continuous structure; but this
is a subject of a forthcoming work.

\section{Oscillation frequencies\label{sec:Results}}

In Fig. \ref{fig:3d_vs_1d} we show the computed oscillation frequencies
in comparison with the observations. The $\hav$ model (black solid
line) clearly exhibits smaller residuals than the 1D solar model,
thereby verifying that it renders the SAR more accurately than the
1D model. We also show a 1D model, in which we expanded the geometrical
depth with the turbulent elevation (dashed line in Fig. \ref{fig:corrections})
and corrected the density for hydrostatic equilibrium. The expansion
improves the 1D model, but it is not enough. A comparison of our $\hav$
model results with \citet{rosenthal_convective_1999} leads to the
same residuals in the oscillation frequencies (see Fig. \ref{fig:rosenthal6})
because the pressure stratifications are almost the same, as we mentioned
above in Sect. \ref{sub:Models}.

\subsection{Corrected structures\label{sub:Corrected-structures}}

\begin{figure*}
\subfloat[\label{fig:without_shift}]{\includegraphics[width=88mm]{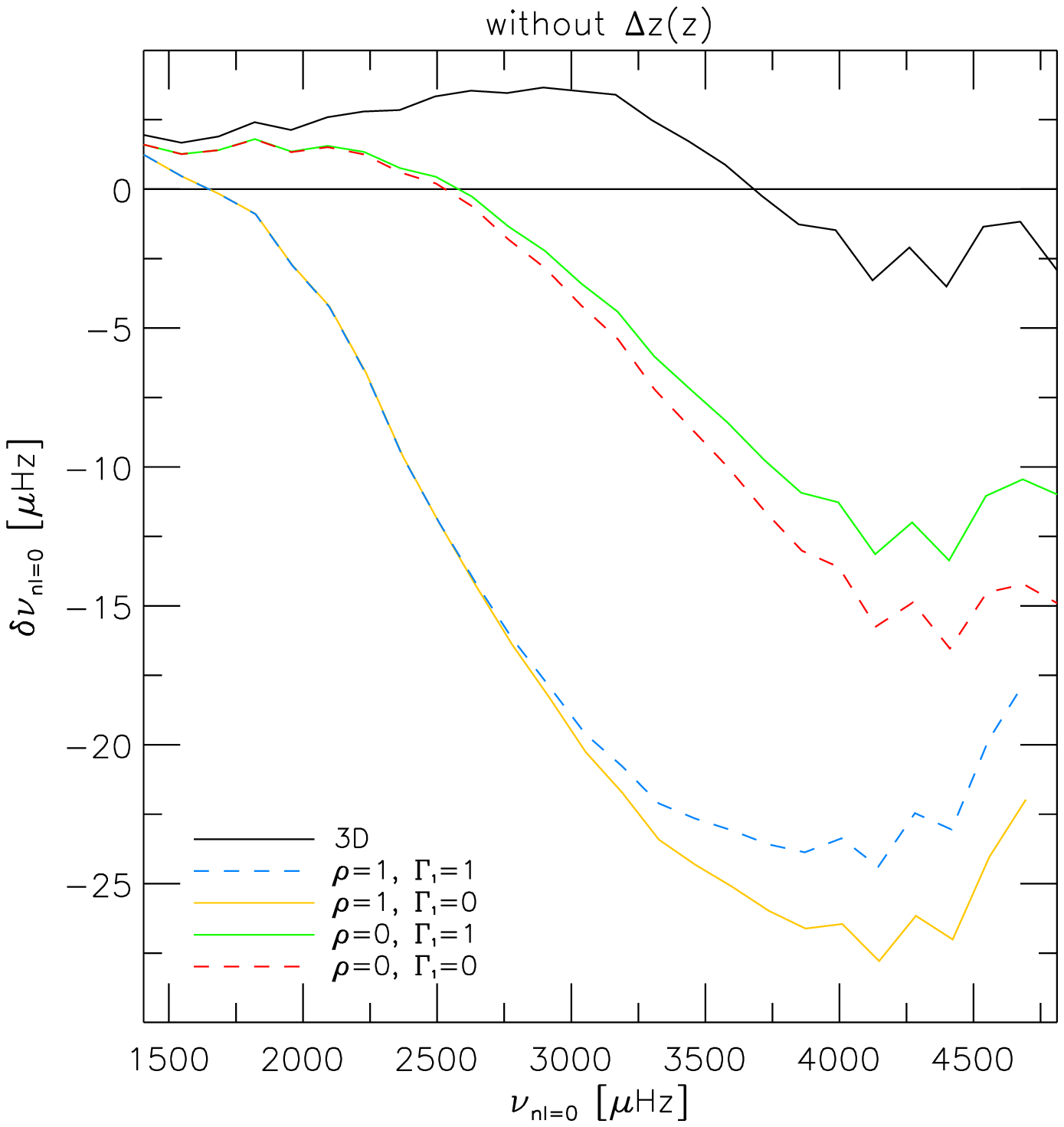}

}\subfloat[\label{fig:with_shift}]{\includegraphics[width=88mm]{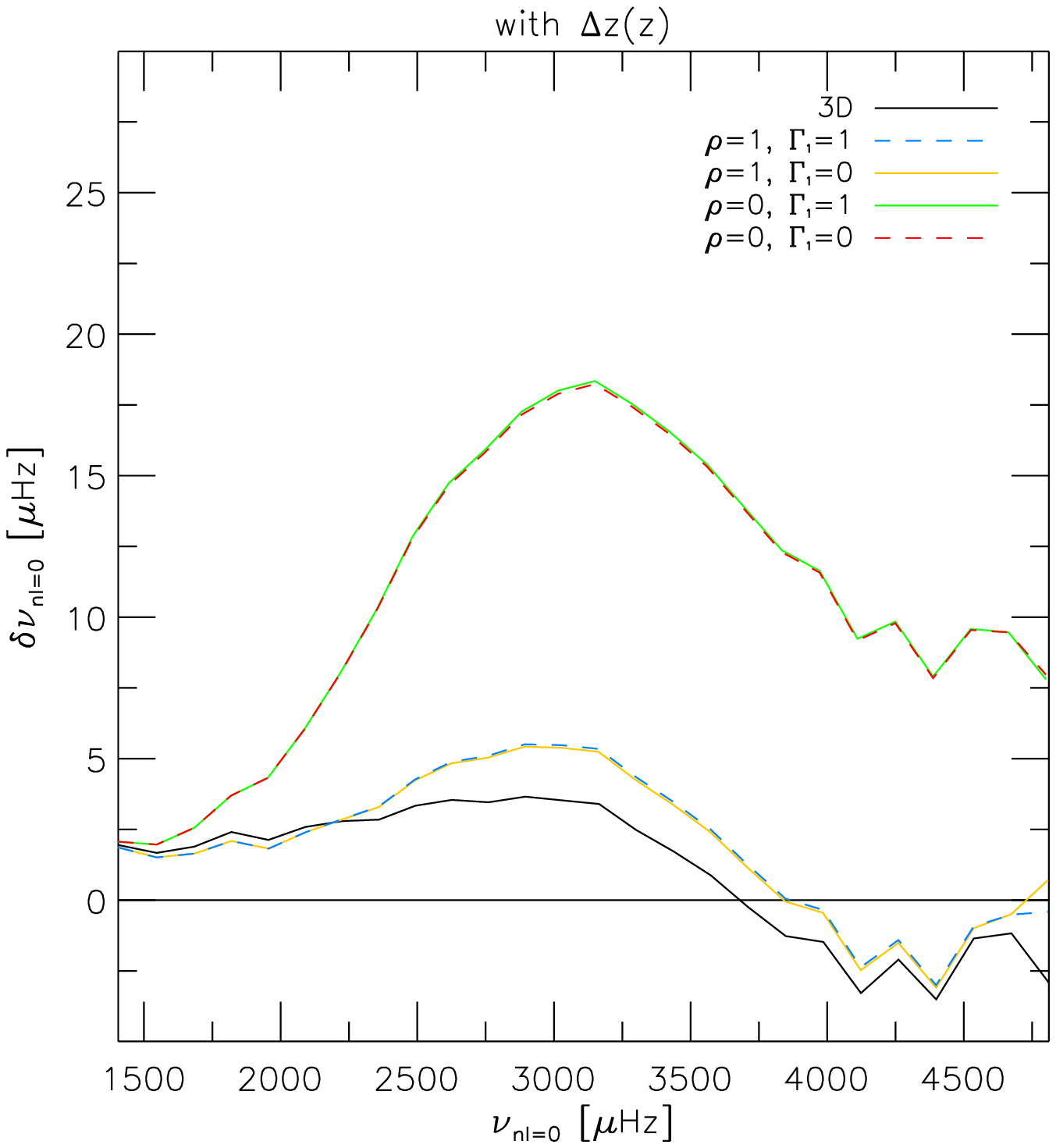}

}

\caption{\textbf{(a)} Similar to Fig. \ref{fig:3d_vs_1d}, but showing 1D models
without any geometrical depth corrections $\Delta z\left(z\right)$,
but with corrections in density and adiabatic exponent. \textbf{(b)}
Similar to Fig. \ref{fig:without_shift}, but here the 1D models include
the geometrical depth corrections $\Delta z\left(z\right)$. Note
the differences in the ordinates. For comparison we also show the
3D model (black solid line) in both panels. See text for more details.
See text for more details.}

\end{figure*}
To determine the actual influences of the corrections in depth, density
and adiabatic exponent, we corrected them individually. First we consider
the models without any depth correction $\Delta z\left(z\right)$
given by Eq. \ref{eq:depth_correction} (see Fig. \ref{fig:without_shift}).
In this panel, all four of the 1D models show large, negative systematic
offsets that reach values of between 12 to $28\,\mu\mathrm{Hz}$ at
the high-frequency end. When we employed only the density correction
to the 1D model (orange solid line), we found that the mismatch was
substantially worsened to $\sim28\,\mu\mathrm{Hz}$. In addition,
adopting the adiabatic exponent alone from the $\hav$ model in the
1D model gives the best match (green solid line), but it does not
improve the mismatch significantly. The change in the adiabatic exponent
affects the oscillation frequencies only slightly with a difference
of $\sim5\mu\mathrm{Hz}$ at higher frequencies that increases above
$\sim3000\,\mu\mathrm{Hz}$ (between $\Gamma_{1}=1$ and $\Gamma_{1}=0$)
compared to the density correction (between $\rho=1$ and $\rho=0$).
This yields a difference of $\sim10\mu\mathrm{Hz}$ (see Fig. \ref{fig:without_shift}).
The adiabatic exponent for the $\hav$ and 1D models is shown in Fig.
\ref{fig:sun_structure} (see blue dashed line in the bottom panel). 

Only by adopting the depth correction $\Delta z\left(z\right)$ given
in Eq. \ref{eq:depth_correction} and the density correction (Eq.
\ref{eq:density_correction}), can we achieve a significant improvement
in the oscillation frequencies in comparison with observations (see
Fig. \ref{fig:with_shift}). The differences in the frequencies between
the 1D model and the observations are reduced by expanding the structure.
It is obvious, however, that the density correction is crucial for
a closer match (blue dashed and orange solid lines; we note that these
two lines overlap in Fig. \ref{fig:with_shift}). The correction of
$\Gamma_{1}$ alone hardly changes anything (blue dashed and green
solid line), which is consistent with the very small differences in
the $\Gamma_{1}$ stratification between the $\hav$ and 1D model
(red lines in bottom panel of Fig. \ref{fig:sun_structure}).

\begin{figure}
\includegraphics[width=88mm]{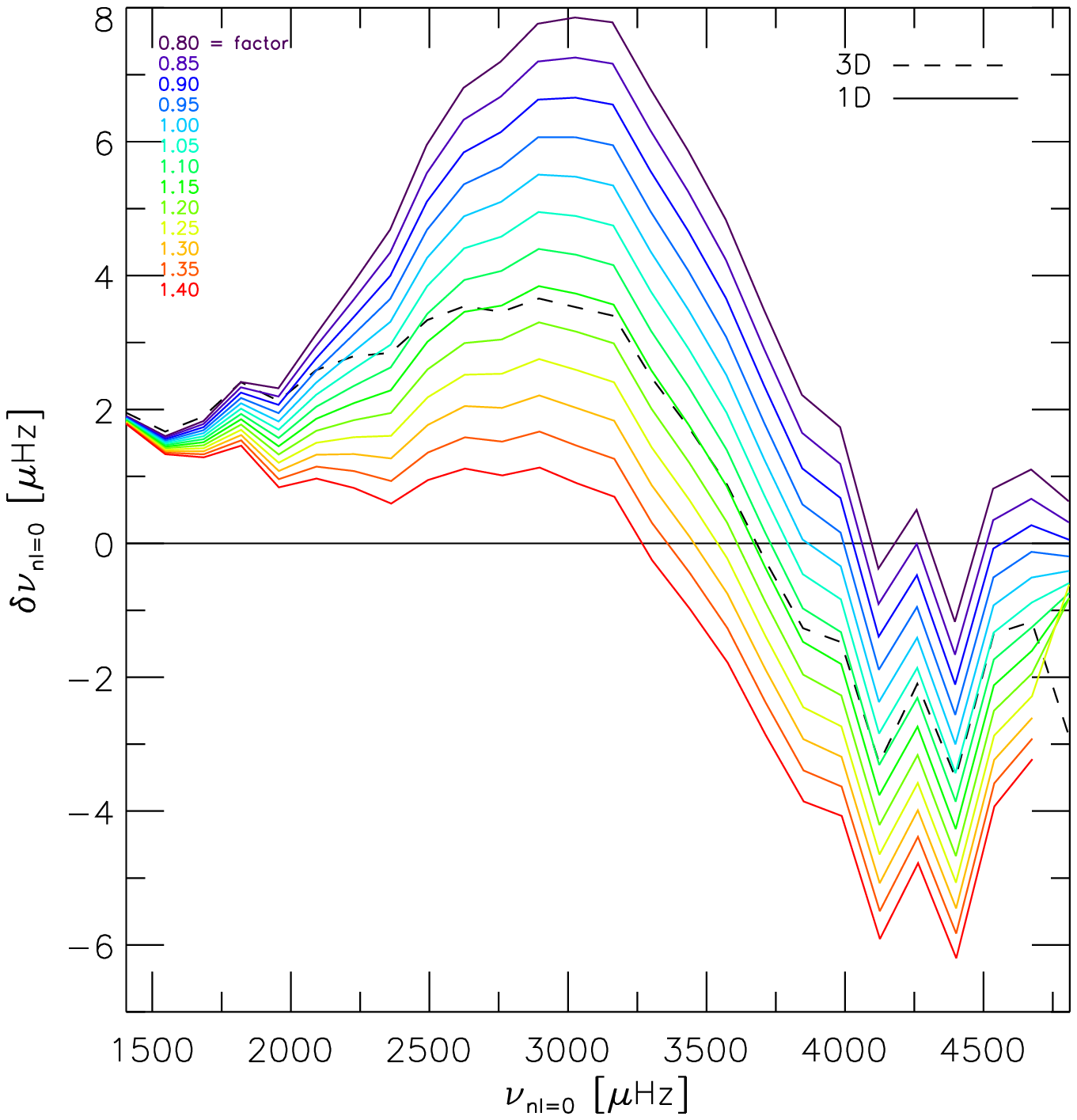}

\includegraphics[width=88mm]{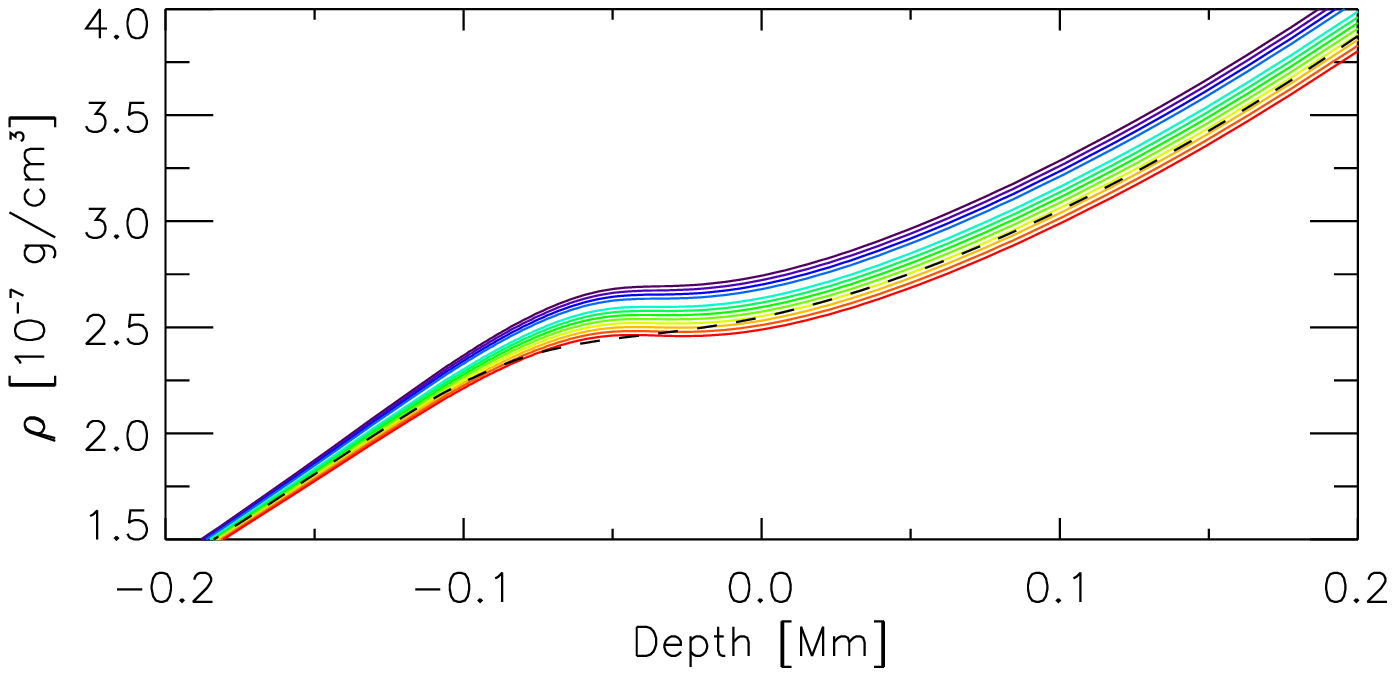}

\caption{\label{fig:density}Top panel: Similar to Fig. \ref{fig:3d_vs_1d},
but showing models with different density corrections. The line with
the density correction of 1.0 is the same as shown in Fig. \ref{fig:with_shift}
(blue dashed line). Bottom panel: The density profiles are shown.
In both panels, the 3D-corrected model is also shown for comparison
(\emph{black dashed line}).}
\end{figure}
The density corrections seem small (see red dotted line in Fig. \ref{fig:sun_structure}),
but their effect on the frequencies is very strong. This illustrates
that the frequencies are very sensitive to the density stratification
in the SAR immediately below the optical surface. The density needs
to be reduced because the missing turbulent pressure is compensated
for by higher densities. We varied the density correction (Eq. \ref{eq:density_correction})
with a constant scaling factor between 0.8 and 1.4, to show the effect
on the p-mode frequency, which is shown in Fig. \ref{fig:density}.
Again, the changes in the frequencies are very strong, indicating
the importance of the density stratification on the oscillations.
A smaller density correction increases the residuals between observations
and theoretical model, in particular around $\sim3000\,\mu\mathrm{Hz}$,
while a larger density correction decreases the residuals, but at
higher frequencies the mismatch is more pronounced. The net effect
of these changes in the residuals is to shift the frequency range
over which the residuals are small from high frequencies for the largest
density corrections to lower frequencies for the smallest density
corrections.

\begin{figure}
\includegraphics[width=88mm]{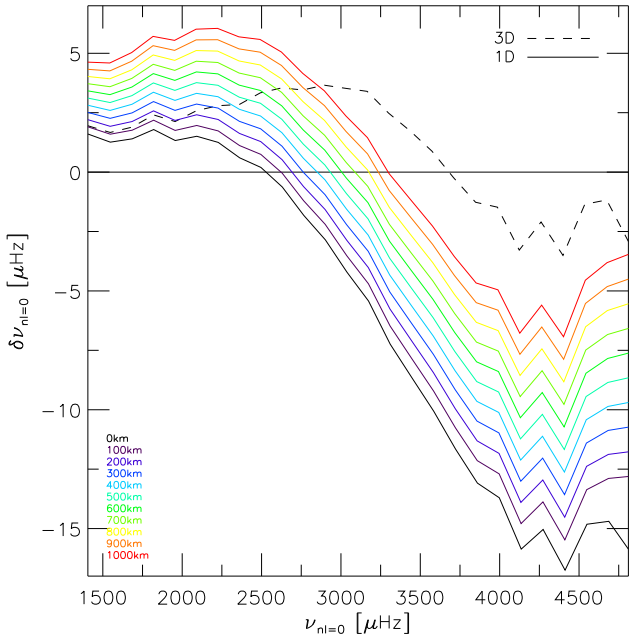}

\caption{\label{fig:rad_exp}Similar to Fig. \ref{fig:3d_vs_1d}, but showing
1D models without any corrections. Instead, only the radii are expanded.
The 3D-corrected model is also shown for comparison (\emph{black dashed
line}).}
\end{figure}
An expansion of the global radius of the model alone will extend the
acoustic cavity and it will also result in increases in the modelled
frequencies. However, such an isolated change to the model will not
resolve the mismatch, since the structures of the surface layers are
unchanged (see Fig. \ref{fig:rad_exp}). The relative differences
between model and observations are shifted to higher values, but the
overall shape is preserved. Therefore, none of the models with expanded
radius can provide a good match to the observations. This illustrates
that the depth-dependent corrections from the $\hav$ model in the
SAR of the solar structure are as important as the expansion of the
radius alone. The expansion of the radius has to be included, but
by itself it is not sufficient to correct for the surface effects.

Instead of correcting the frequencies for the surface-effects, we
can also directly correct the surface layers of a 1D model. As shown
in Fig. \ref{fig:with_shift}, a depth-dependent expansion of the
geometrical depth and a correction of the density are necessary steps.
Furthermore, from asteroseismic observations, we can infer the correct
structure of stellar structure model-independently models by finding
the right corrections around the critical surface layers.

\subsection{Different averages\label{sub:Averages}}

\begin{figure}
\includegraphics[width=88mm]{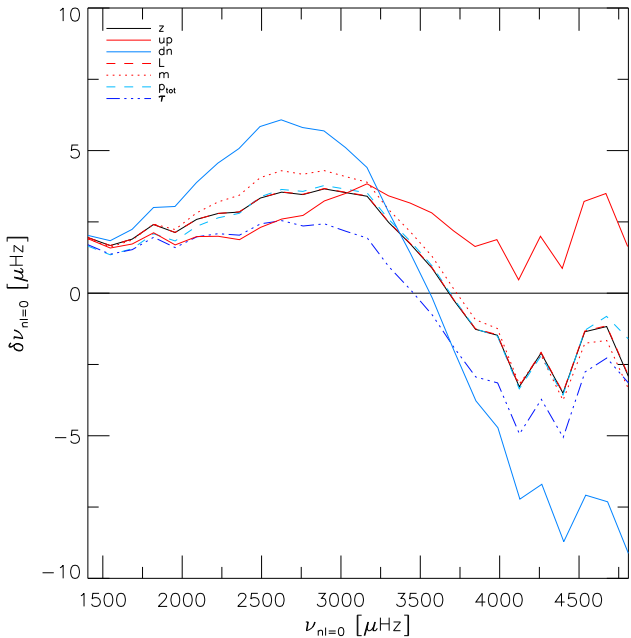}

\caption{\label{fig:averages}Similar to Fig. \ref{fig:3d_vs_1d}, but showing
models with different $\hav$ averages.}
\end{figure}
The question arises about the effect of different averaging methods
of the 3D model structure on the p-mode frequencies. In Fig. \ref{fig:sun_averages}
we compared the radial profiles of pressure, density, and adiabatic
exponent that we obtained using seven different averaging methods
of the 3D model structure. We found above in Sect. \ref{sub:Averaging}
that most of the different averages depict only small changes, which
also results in only small changes of the oscillation frequencies
as shown in Fig. \ref{fig:averages}. The geometrical averages for
the up- and downflows exhibit the largest difference in the density
and adiabatic exponent immediately below the optical surface. The
upflows depict lower densities, because the hotter granules are lighter
than the turbulent downdrafts in the intergranular lane. The $\hav$
model for the downflows leads to the highest mismatches with observations
compared to the other $\hav$ model. On the other hand, the $\hav$
model of the upflows leads to the best agreement with observations,
since the residuals are almost a straight line. In particular, the
agreement at higher frequencies is much better, which is mainly due
to the adiabatic exponent. One reason for this result might be the
adiabatic nature of the frequency calculations, since the averages
over the upflows are closer to the adiabatic stratification. The pseudo-Lagrangian
averages are indistinguishable from the plain geometrical averages.
The averages on the acoustic depth are slightly closer to observations
below $\sim3000\,\mu\mathrm{Hz}$, but in higher frequencies the mismatch
is larger. The averages on constant layers of column mass density
and total pressure are about the same as on geometrical depth.

\subsection{Chemical composition\label{sub:Abundances}}

\begin{figure}
\includegraphics[width=88mm]{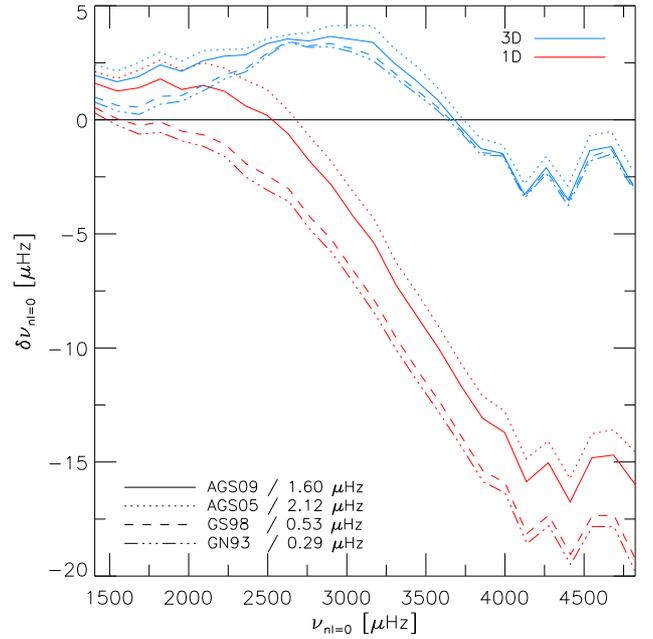}

\includegraphics[width=88mm]{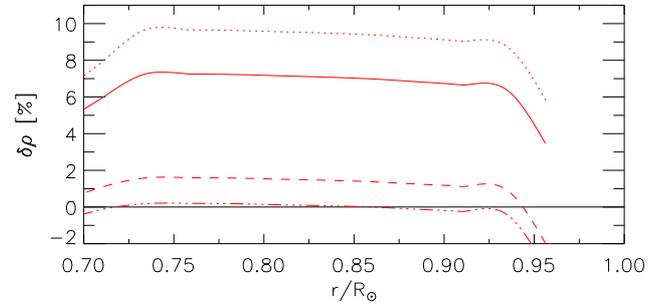}

\caption{\label{fig:abund}Top panel: Similar to Fig. \ref{fig:3d_vs_1d},
but showing models with different abundances. Shown are uncorrected
1D models (\emph{red lines}) and 3D-corrected models (\emph{blue lines}).
The latter are the different 1D models, but appended with the same
3D model, which is why they converge towards higher frequencies and
differ towards lower frequencies. We also note the low-frequency mismatch
for the individual abundances. Bottom panel: The differences in density
compared to inferred $\rho$ from observations \citep{2009ApJ...699.1403B}. }
\end{figure}
Next we consider the influence of the chemical composition on the
oscillations frequency. In Fig. \ref{fig:abund} we show frequencies
computed with solar 1D models assuming solar chemical compositions
different from our standard composition \citep[AGS09]{asplund_chemical_2009},
while the abundances in the $\hav$ model were fixed (the difference
would be only minor). We considered the solar chemical compositions
by \citet[AGS05]{2005ASPC..336...25A}, \citet[GS98]{1998SSRv...85..161G},
and \citet[GN93]{1993oee..conf...15G}. The modes at low frequency
are affected by a change in chemical composition. For chemical compositions
assuming higher metallicity, the low frequencies match the solar observations
better. With the most metal-poor composition AGS05 we find a mismatch
of $2.12\,\mu\mathrm{Hz}$ at the low-frequency end ($1500\,\mu\mathrm{Hz}$),
while for most of the metal-rich composition GN93 we find the best
match with much lower residuals with $0.3\,\mu\mathrm{Hz}$. The different
chemical compositions change the structure in the solar models. The
models with higher metallicity match the interior better, which is
also known for the inferred sound speed and density in the convection
zone (see Fig. \ref{fig:abund}).

\subsection{Magnetic field\label{sub:Magnetic-field}}

\begin{figure}
\includegraphics[width=88mm]{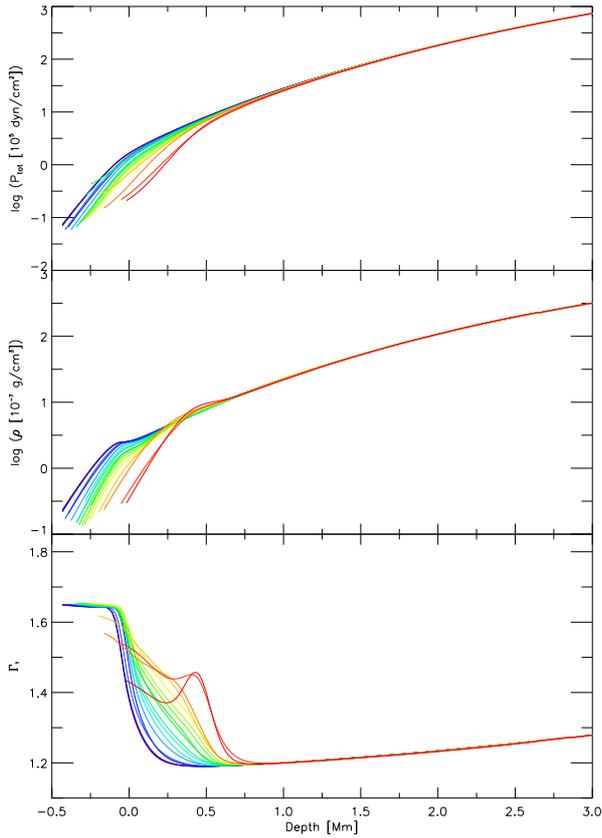}

\caption{\label{fig:magnetic_struct}Comparison of total pressure, density
and adiabatic exponent for $\hav$ models with different magnetic
field strengths. }
\end{figure}
\begin{figure}
\includegraphics[width=88mm]{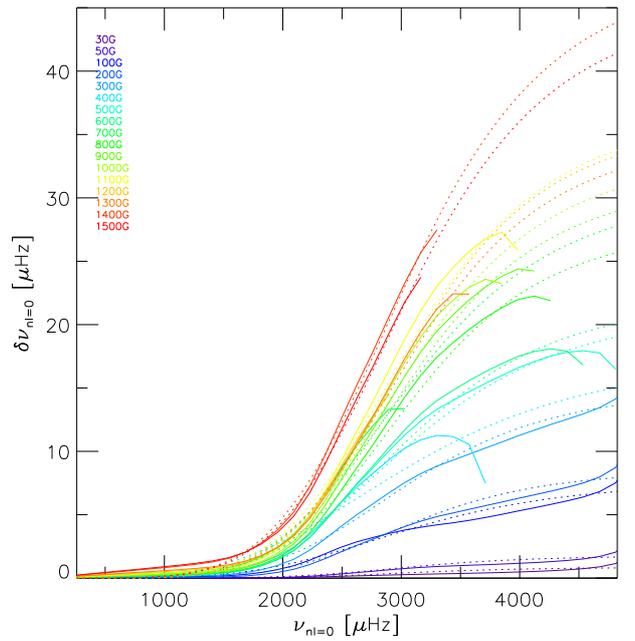}

\caption{\label{fig:magnetic}Frequency differences between $\hav$ models
with different magnetic field strengths compared to non-magnetic case,
i.e. $\delta\nu=\nu_{X\mathrm{G}}-\nu_{0\mathrm{G}}$. We also performed
a modified Lorentzian fit (\emph{dotted lines}).}
\end{figure}
\begin{figure}
\includegraphics[width=88mm]{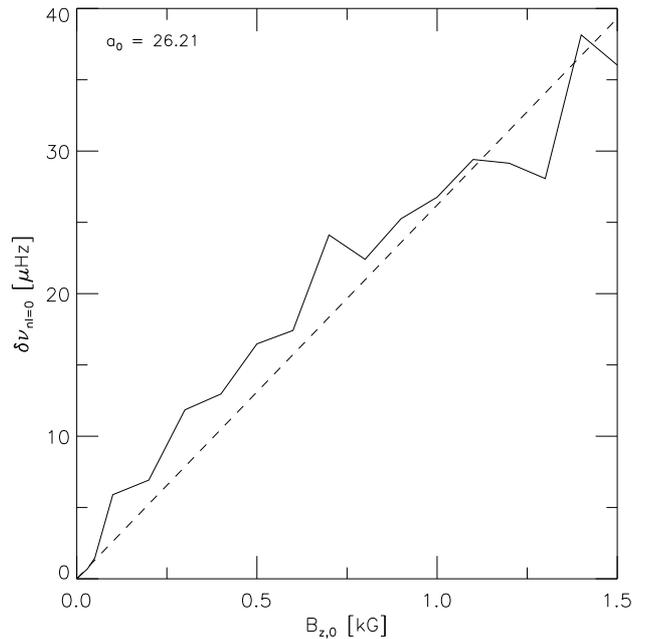}

\caption{\label{fig:magnetic_shift}Frequency shifts determined at $\nu_{nl}=4000\,\mu\mathrm{Hz}$
from the modified Lorentzian fits (Fig. \ref{fig:magnetic}), shown
against the magnetic field strength (\emph{black solid line}). We
also show the linear fit (\emph{dashed line}).}
\end{figure}
\begin{table}
\caption{\label{tab:hyp_fit}Coefficients $\alpha$ and $\beta$ for the modified
Lorentzian function (Eq. \ref{eq:modified_lorentzian}) shown in Fig.
\ref{fig:magnetic}.}

\begin{tabular}{rccrcc}\hline\hline
$B_{z,0}$ & $\alpha$ & $\beta$ & $B_{z,0}$ & $\alpha$ & $\beta$\\
\hline
50G & 0.278 & 5.940 & 800G & 10.014 & 4.888 \\
100G & 0.576 & 5.885 & 900G & 9.103 & 5.290 \\
200G & 2.527 & 4.369 & 1000G & 10.242 & 5.322 \\
300G & 2.799 & 5.384 & 1100G & 10.893 & 5.254 \\
400G & 4.912 & 4.917 & 1200G & 12.018 & 5.180 \\
500G & 5.785 & 3.755 & 1300G & 11.711 & 5.503 \\
600G & 6.930 & 4.670 & 1400G & 11.391 & 5.309 \\
700G & 7.249 & 4.848 & 1500G & 15.675 & 5.076 \\
\hline\end{tabular}
\end{table}
We also computed solar 3D MHD models with vertical magnetic field
strengths of up to 1500 kG. We included monolithic vertical fields
in initially hydrodynamic models and appointed the same vertical magnetic
field to the gas entering the numerical box at the bottom. Then, we
evolved the simulations until the MHD models reached the quasi-stationary
state. From the subsequent models we determined spatial and temporal
averages. We truncated the initial hydrodynamic solar model slightly
at the top and we reduced the numerical resolution to $120^{3}$ to
keep computational costs low. The magnetic inhibition of the flow
increases with the magnetic field strength through the coupling of
the velocity field with the magnetic field through the Lorentz force
in the SAR. This leads to significant changes in the stratification
of the solar model. Owing to the magnetic inhibition of convection,
the mean temperature, density, and pressure are reduced significantly
in the surface layers (see Fig. \ref{fig:magnetic_struct}). Furthermore,
the turbulent pressure is reduced, and the magnetic pressure increasingly
evacuates the plasma. In this way the optical surface is depressed
by 400 km in the model with 1.5 kG, which affects the frequencies
by shortening the acoustic cavity, which in turn increases the oscillation
frequencies. We appended the magnetic $\hav$ models to the 1D solar
model to test the influence of the magnetic field on the oscillation
frequencies. We show in Fig. \ref{fig:magnetic} the corresponding
comparisons with the non-magnetic model. As a consequence of the depression
of the optical surface, the frequencies are enhanced with increasing
field strength. At high field strength with 1.5 kG, the p-mode frequencies
are increased by almost $\sim45\,\mu\mathrm{Hz}$ at the high-frequency
end. We also determined the relation between magnetic field strength
and the frequency shift, $\delta\nu_{nl}\propto B_{z}$ (see Figs.
\ref{fig:magnetic} and \ref{fig:magnetic_shift}). We fitted the
relative differences with the modified Lorentzian function
\begin{eqnarray}
\delta\nu/\nu_{\mathrm{max}} & = & \alpha\left(1-1/\left[1+(\nu/\nu_{\mathrm{max}})^{\beta}\right]\right)\label{eq:modified_lorentzian}
\end{eqnarray}
with $\nu_{\mathrm{max}}=3090\,\mu\mathrm{Hz}$ \citep[see][]{ball_new_2014}
and determined the shift at $\nu_{nl}=4000\,\mu\mathrm{Hz}$. We found
the following response relation for the magnetic field 
\begin{eqnarray}
\delta\nu_{nl} & = & 26.21B_{z},\label{eq:response}
\end{eqnarray}
where the vertical magnetic field strength is given in units of kG
(see Fig. \ref{fig:magnetic_shift}). We note that the frequency shift
considered at $\nu_{\mathrm{max}}=3090\,\mu\mathrm{Hz}$ would instead
yield a smaller slope of $16.57$. The changes in oscillation frequencies
are due to both the reduction of the radius and the altered stratification
of the independent variables. The latter has the stronger effect,
in particular for the high-frequency shifts, since we tested this
by keeping the radius unchanged, which led to very similar results.
Furthermore, for the higher angular degrees modes of $l=1,2$ and
$3$, we find very similar slopes with $25.80,\,25.92$, and $26.01$.
In Table \ref{tab:hyp_fit}, we list the coefficients of the modified
Lorentzian function (Eq. \ref{eq:modified_lorentzian}) that result
from the fitting for $l=0$.

\begin{figure*}
\subfloat[\label{fig:mag_mdi_hist}]{\includegraphics[width=88mm]{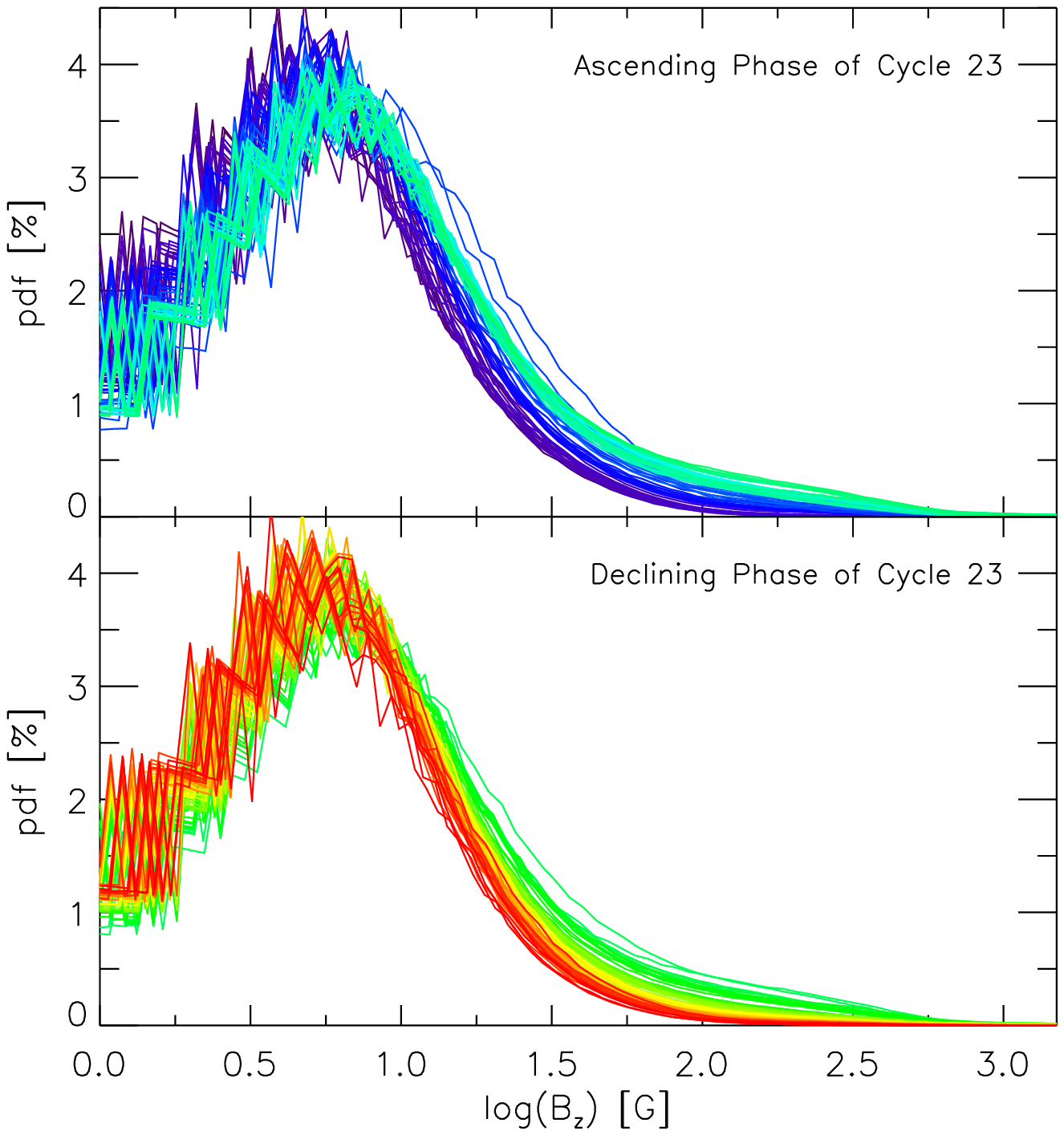}

}\subfloat[\label{fig:mag_total_shift}]{\includegraphics[width=88mm]{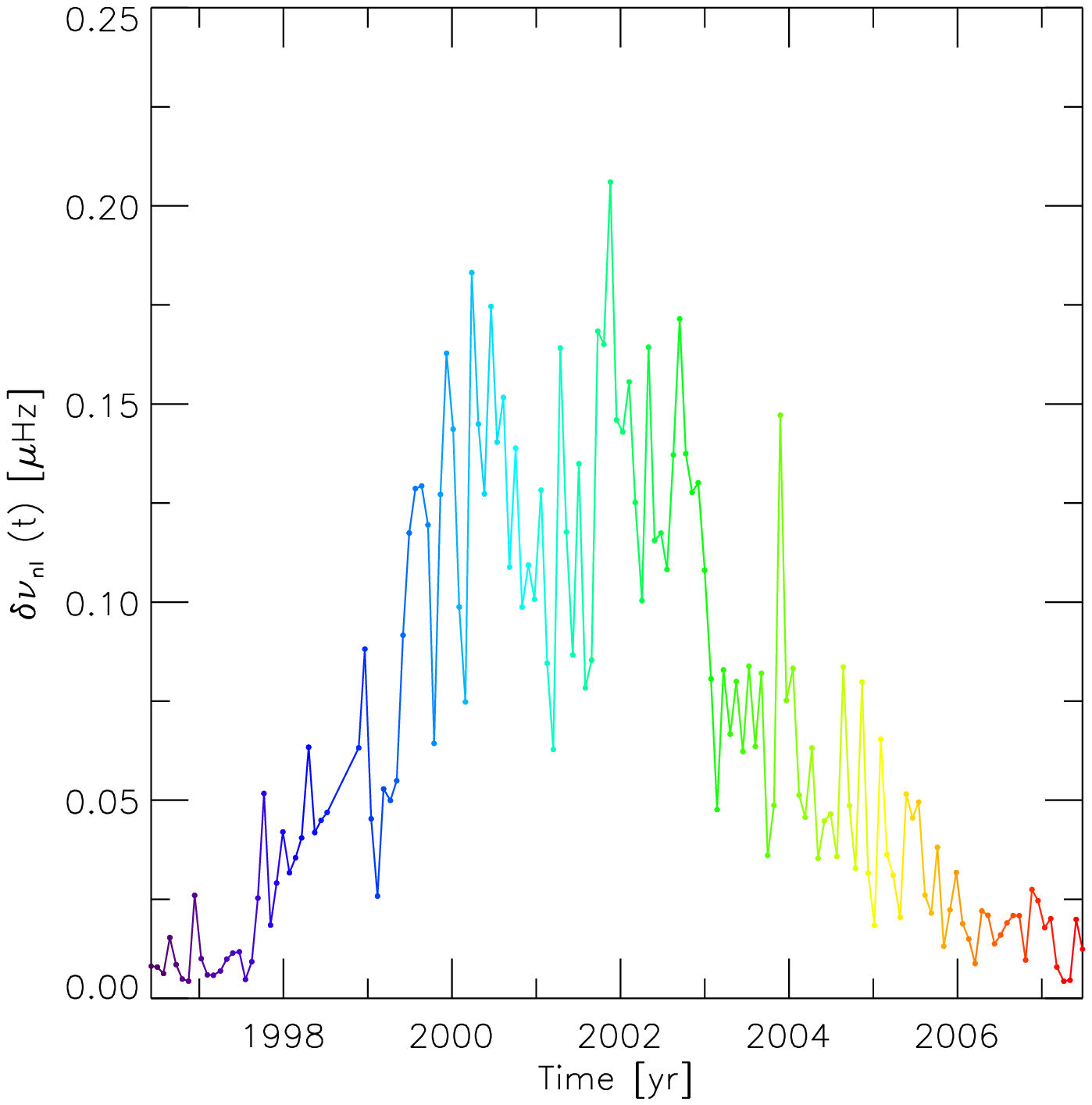}

}

\caption{\textbf{(a)} Histogram of magnetic field strength determined from
synoptic SOHO/MID maps for the ascending and declining phases (\emph{top}
and \emph{bottom} panel, respectively). \textbf{(b)} The integrated
frequency shift, $\delta v_{nl}\left(t\right)$ (Eq. \ref{eq:total_shift}),
vs. time for solar cycle 23. Both figures have the same colour-coding
for time.}
\end{figure*}

To estimate the global impact on the frequency, we determined the
probability distribution function of the magnetic field on the solar
surface, $f_{B_{z}}$, during the different phases of the solar cycle.
We determined $f_{B_{z}}$ for solar cycle 23 (Carrington rotation
numbers from 1909 till 2057, i.e. between the years 1996 and 2008)
from synoptic (radially corrected) magnetograms observed by SOHO/MDI\footnote{Retrieved from \url{http://soi.stanford.edu}.}
(see Fig. \ref{fig:mag_mdi_hist}). We computed the histograms for
100 bins from the logarithm of the absolute magnetic field strength
values. As expected, during the ascending phase of cycle 23, the probability
of the magnetic field strength increases at higher field strength,
which reduces the probabilities at lower $B_{z}$ (the total probability
is unity). The opposite is true for the declining phase. 

Then, with the linear response function (Eq. \ref{eq:response}),
we can determine the total shift by integrating the probability distribution
function of the surface magnetic field, $f_{B_{z}}$, over the magnetic
field strength, 
\begin{eqnarray}
\delta v_{nl}\left(t\right) & = & \frac{1}{B_{z,\mathrm{max}}}\int_{0}^{B_{z,\mathrm{max}}}\delta\nu_{n0}^{*}(B_{z},t)f_{B_{z}}(B_{z},t)\,dB_{z}.\label{eq:total_shift}
\end{eqnarray}
We calculated the total shift (Eq. \ref{eq:total_shift}) for solar
cycle 23, which is shown in Fig. \ref{fig:mag_total_shift}. The total
shift is largest during the maximum (2002), leading to a shift of
$\sim0.2\,\mu\mathrm{Hz}$. \citet{2004MNRAS.352.1102C} considered
the frequency shifts for different phases during solar cycles 22 and
23. For the low-angular degree $l=0/2$ at the solar activity maximum
they found a mean shift around $\sim0.2/0.4\,\mu\mathrm{Hz}$. We
found very good agreement with our theoretical results. However, we
did not find a strong dependence on angular degree mode. From 3D MHD
models for stars other than the Sun, the linear response functions,
$\delta\nu_{nl}^{*}$, can be predicted. This opens the possibility
of inferring their magnetic field distribution function, $f_{B_{z}}$,
at their surface, by comparing and matching $f_{B_{z}}$ to observations
of their stellar cycles.

\section{Conclusions\label{sec:Conclusions}}

The turbulent expansion that elevates the depth-scale and reduces
the density stratification is very important for matching the observed
solar oscillation frequencies more accurately. On the other hand,
the differences in the stratification of the adiabatic exponent are
often very small, therefore, the their effects on on the oscillation
frequencies are accordingly less important. Furthermore, instead of
correcting for the frequencies, the surface of the solar model can
be corrected for by expanding the the geometrical depth scale in a
depth-dependent way and by reducing the density by the turbulent pressure.
This leads to very similar results, as achieved with a $\hav$ model
appended to a 1D model. Considering alternative reference depth scales
for determining the spatial averages for the 3D model leads to very
similar results, as achieved with averages over geometrical depth.
We also found that 1D solar models with higher metallicity result
in models that match the low-frequency part that probe the interior
better. Finally, we found that strong magnetic fields have a distinct
influence by predominantly increasing the high-frequency range and
that the linear response is able to reproduce the solar activity cycles
properly.

\begin{acknowledgements}
This work was supported by a research grant (VKR023406) from VILLUM
FONDEN.
\end{acknowledgements}

\bibliographystyle{aa}
\bibliography{papers}

\end{document}